 \documentclass[final,5p,times,twocolumn]{elsarticle}

\usepackage{amssymb}
\usepackage{lipsum}

\usepackage[usenames]{color}
\usepackage{colortbl}

\journal{Annals of Physics}

\begin{document}

\begin{frontmatter}



\title{Observable strong field effects of extra spacetime dimension in the braneworld black hole}


\author[first,second]{K.K. Nandi}
\author[first]{R.N. Izmailov}
\author[first]{R.Kh. Karimov}
\author[third]{A.A. Potapov}

\affiliation[first]{organization={Zel'dovich International Center for Astrophysics, M. Akmullah Bashkir State Pedagogical University},
            addressline={3A, October Revolution Street}, 
            city={Ufa},
            postcode={450008}, 
            state={RB},
            country={Russia}}
\affiliation[second]{organization={High Energy Cosmic Ray Research Center, University of North Bengal},
            city={Siliguri},
            postcode={734013}, 
            state={WB},
            country={Bharat}}
\affiliation[third]{organization={Ufa University of Science \& Technology, Sterlitamak Campus},
            addressline={Lenin Avenue, 49}, 
            city={Sterlitamak},
            postcode={453103}, 
            state={RB},
            country={Russia}}

\begin{abstract}
Inspired by the string theory, the braneworld picture introduces extra dimensions beyond the four that may have observable non-trivial effects in short distance (strong field) gravity experiments. A case in point is the Randall-Sundrum braneworld picture that projects the $5d$ bulk Weyl tensor onto the $3d$ brane providing a stress tensor in the effective Einstein field equations on the brane. Dadhich, Maartens, Papadopoulos and Rezania (DMPR) derived an exact braneworld black hole solution of the brane vacuum field equations. The solution formally resembles that of Reissner-Nordstr\"{o}m but is physically different from it since the "tidal charge" $\Upsilon$ in the solution is not the electric charge but an imprint from the fifth dimension allowing both signs in the power law modification $\pm \frac{\Upsilon ^{2}}{r^{2}}$ to the Schwarzschild metric $\left( \Upsilon = 0\right)$. The corresponding black holes are designated as DMPR$\pm$. We study here the effect of $\Upsilon$ on strong field lensing observables and compare in the eikonal limit the ring down quasinormal mode (QNM) frequencies of DMPR$-$ with those of DMPR$+$, the two variants of tidal charge modified Schwarzschild black hole ($\Upsilon = 0$). It turns out that the tidal charge can significantly modify the Schwarzschild lensing observables and QNM frequencies. In particular, we find that the Pretorius-Khurana critical exponent $\gamma$ of circular null orbits in the DMPR$-$ black hole has a lower value than that for the Schwarzschild black hole, which indicates a stronger Lyapunov instability suggesting that the accretion disks of DMPR$-$ black holes would appear brighter. The case of the SgrA* black hole is considered for a possible constraint on $\Upsilon$ from the EHT observation of its shadow size.
\end{abstract}







\end{frontmatter}




\section{Introduction}
\label{introduction}

The existence of the extra dimensions, such as the one introduced by the fifth dimension in the braneworld scenario or in string theory, may have non--trivial observable effects in gravity experiments in the short distance limit. Such experiments should necessarily involve strong field effects of black holes arising from the collapse of matter localized on the visible $3$-brane in the $(3+1+1)d$ bulk in the Randall-Sundrum braneworld scenario \cite{Randall:1999}. Conformally symmetric vacuum solutions have been found in the braneworld model by Mak and Harko \cite{Harko:2005}. New properties of the Shiromizu et al \cite{Shiromizu:2000} formalism have been recently found by Culeto \cite{Culetu:2021}. This model has been applied to the galactic halo for exploring many new features \cite{Nandi:2009}. Timing effects in the Kerr-Sen string metric have been studied in \cite{Izmailov:2020}. When matter fields in the $3$-brane collapse to form a static spherically symmetric black hole, it should yield the observable effects through the modifications of Schwarzschild gravity without warping, when the $4d$ cosmological constant $\Lambda _{4}$ is zero. Such a collapse was studied in \cite{Chamblin:2000} which yielded an infinite `black string' solution intersecting the $3$-brane in a Schwarzschild solution. The fifth dimension in the braneworld could be non-compact and infinitely large \cite{Arkani:1998,Rubakov:1983} in which gravity is assumed to be freely propagating. However, there is a caveat here. As exemplified by Gregory et al \cite{Gregory:2000}, the picture that direct manifestations of large extra dimensions can occur only at short distance scales, while the long-distance physics is effectively four-dimensional, is not universally true.

However, in the Randall-Sundrum scenario \cite{Randall:1999}, the caveat does not apply since the manifestation of fifth dimension occur only in the short-distance limit on the brane, where the lowest order correction to the Schwarzschild potential is proportional to $\frac{1}{r^{2}}$ which dominates the $\frac{1}{r}$ term. This feature is revealed from the \textit{exact} black hole solution on the brane found by Dadhich, Maartens, Papadopoulos and Rezania (hereinafter DMPR) \cite{Dadhich:2000}, which thus describes the strong gravity regime on the brane (see also \cite{Casadio:2002}). They found that the solution of the effective Einstein equations on the brane resembles Reissner-Nordstr\"{o}m solution without the usual electric charge but instead a "tidal charge" $\Upsilon$ (TC) as an imprint from the fifth dimension on the brane. The effect of this imprint is to modify the Schwarzschild black hole by introducing power law corrections $\pm \frac{\Upsilon^{2}}{r^{2}}$ to it. We shall call the metric with $+ \frac{\Upsilon^{2}}{r^{2}}$ the RN-type (abbreviated as DMPR$+$) and the one with $- \frac{\Upsilon^{2}}{r^{2}}$ the TC-type black hole (abbreviated as DMPR$-$). Of particular interest is the latter, when the metric describes a single horizon with a radius larger than that of the Schwarzschild thus showing lower temperature and greater entropy. Further it was shown that the negative tidal charge ($- \Upsilon ^{2}$) may provide a mechanism for which its rotation parameter exceeds its mass \textit{without} creating a naked singularity \cite{Aliev:2005}. All these phenomena are not admissible in the spacetime of general relativistic black holes. Therefore, a study of other new non-general relativistic phenomena encapsulated in the strong gravity regime of the DMPR$\pm$ black hole should be quite desirable. The astrophysically observable imprints of extra dimension have not yet been adequately explored in the literature, to our knowledge.

The purpose of the paper is to fill this gap by assuming the DMPR$\pm$ black holes as gravitational lenses and studying the most powerful effects such as embodied in the strong field lensing observables and the emission of QNM frequencies of the DMPR$\pm$ braneworld black holes. These are observable signatures of the extra dimension that could be distinguishable from the general relativistic Schwarzschild black hole.

In Sec.2, we briefly outline the field equations on the brane including the exact black hole solution found by DMPR. In Sec.3, we shall apply the methodology of Bozza \cite{Bozza:2002} for finding the strong field lensing signatures of the lens DMPR black holes representing the supermassive black hole SgrA* existing in the center of our galaxy. In Sec.4, we calculate the QNM modes, Lyapunov and Pretorius-Khurana exponents. A prime candidate for observing the $\Upsilon-$effect is the angular diameter of the shadow of SgrA*. Sec.5 briefly discusses the issues related to its measurement and an upper limit on $\Upsilon$ is proposed. Sec.6 concludes the work. We choose units in which $G=1$, $c=1$ unless the symbols are specifically restored.

\section{Braneworld field equations and the DMPR solutions}
\label{sec2}

The original Randall-Sundrum model \cite{Randall:1999} assumes the bulk to be anti-de Sitter. Shiromizu et al \cite{Shiromizu:2000} developed a geometric approach of the model that satisfies the $5-$dimensional Einstein equations with cosmological constant. We shall only briefly outline for completeness the braneworld field equations and the DMPR solution, while the details can be found in \cite{Dadhich:2000}. The relevant field equations can be written as modified Einstein equations with the new terms indicating projected bulk effects onto the brane:
\begin{equation}
G_{\mu\nu} = -\Lambda g_{\mu \nu }+\kappa ^{2}T_{\mu \nu }+\widetilde{\kappa 
}^{4}S_{\mu \nu }-\mathcal{E}_{\mu \nu },
\end{equation}
where $\Lambda $ is the cosmological constant on the brane, $\kappa^{2} = \frac{8\pi }{M_{p}^{2}}$, $\widetilde{\kappa}^{2} = \frac{8\pi}{\widetilde{M}_{p}^{3}}$, $M_{p}$ and $\widetilde{M}_{p}$ are the effective fundamental Planck masses on the brane and the bulk respectively, $T_{\mu \nu}$ is the usual energy-momentum tensor, $S_{\mu \nu }$ is the squared energy-momentum tensor, $\mathcal{E}_{\mu \nu }$ is the limit on the $3-$brane of the projected bulk Weyl tensor $\widetilde{C}_{ACBD}$ such that
\begin{equation}
\mathcal{E}_{AB}=\widetilde{C}_{ACBD}n^{C}n^{D}.
\end{equation}%
In the brane vacuum, $\Lambda _{4}=0$ corresponding to a negative cosmological constant $\widetilde{\Lambda }_{5}$ in the bulk and $T_{\mu\nu} = S_{\mu\nu} = 0$, so that the closed system of braneworld field equations, due to the symmetries of the Weyl tensor, are given by 
\begin{eqnarray}
R_{\mu \nu } &=&-\mathcal{E}_{\mu \nu }, \\
R_{\mu }^{\mu } &=&0=\mathcal{E}_{\mu }^{\mu }, \\
\nabla ^{\mu }\mathcal{E}_{\mu \nu } &=&0.
\end{eqnarray}%
Even though $\mathcal{E}_{\mu \nu }$ is trace free, the field equations are physically not the same as the Einstein-Maxwell equations since $\mathcal{E}_{\mu\nu}$ is a projection of the bulk Weyl tensor and not the usual Maxwell field tensor $F_{\mu\nu}$.

A vacuum black hole solution of the effective field equations on the brane is given by the ansatz
\begin{equation}
d\tau ^{2}=-A(r)dt^{2}+B(r)dr^{2}+C(r)\left( d\theta ^{2}+\sin ^{2}\theta
d\varphi ^{2}\right).
\end{equation}%
Due to the the projection of free gravitational field effects in the bulk onto the brane, the corrected Schwarzschild potential $-\frac{M}{M_{p}^{2}r}$ is given by%
\begin{equation}
\Phi =-\frac{M}{M_{p}^{2}r}+\frac{\Upsilon ^{2}}{2r^{2}},
\end{equation}%
where $\Upsilon ^{2}>0$ is the squared tidal charge parameter with dimension $\left[L\right] ^{2}$\footnote{%
In place of $Q$ in \textit{Eq.(1)} of \cite{Dadhich:2000}, we have intentionally chosen the constant to be $\Upsilon^{2}$ in our corresponding Eqs.(11,14) only for the sake of notational consistency with the standard form of Reissner-Nordstr\"{o}m metric in the literature. No loss of generality is involved here.}.

DMPR obtained the exact black hole solution of the effective field Eqs.(3)-(5) on the brane that is given by the induced metric \cite{Dadhich:2000} (see also \cite{Casadio:2002}):
\begin{eqnarray}
A(r) &=& \frac{1}{B(r)} = 1-\left( \frac{2M}{M_{p}^{2}}\right) \frac{1}{r} + \frac{\Upsilon^{2}}{r^{2}}, \\
C(r) &=&r^{2}.
\end{eqnarray}%
It is convenient to define the measure of distance scaled by the Schwarzschild radius 
\begin{equation}
r_{s} = 2M = 1
\end{equation}%
(in units $M_{p}=1$) so that the braneworld metric with tidal charge now assumes the form of what we call the DMPR$+$ metric:
\begin{eqnarray}
A_{+}(r) &=&\frac{1}{B_{+}(r)}=1-\frac{1}{r}+\frac{\Upsilon ^{2}}{r^{2}}, \\
C(r) &=&r^{2}.
\end{eqnarray}%
The solution (8) shows two types of braneworld black holes with two different $4d$ horizon structures depending on the sign before $\Upsilon^{2}$. The DMPR$+$ metric is the general relativistic analogue of the Reissner-Nordstr\"{o}m metric, when the sign before $\Upsilon^{2}$ is positive, as in (8), which has \textit{two} horizon radii given by 
\begin{equation}
r_{h}^{\pm }=\frac{1}{2}\left[ 1\pm \sqrt{1-4\Upsilon ^{2}}\right] \leq 1
\end{equation}%
already introducing a limit on the tidal charge $\Upsilon \leq \frac{1}{2}$.

The more interesting case is to allow a negative sign before $\Upsilon^{2}$ as in the metric (14) below, which is also an equally valid solution of the braneworld field Eqs.(3)-(5). To achieve that, what we need to do is only to replace $\Upsilon \rightarrow i\Upsilon $ in (8) obtaining a \textit{non-general relativistic} tidal charge DMPR black hole on the brane. Indeed, we shall see later that this replacement can have observable implications that are considerably distinct from that of the analogue Reissner-Nordstr\"{o}m spacetime (8). The metric (8) now assumes the form which we call the DMPR$-$ metric: 
\begin{eqnarray}
A_{-}(r) &=&\frac{1}{B_{-}(r)}=1-\frac{1}{r}-\frac{\Upsilon ^{2}}{r^{2}}, \\
C(r) &=&r^{2}.
\end{eqnarray}%
Note that $\Upsilon^{2} > 0$, in both metrics (11) and (14). Metric (14) yields a \textit{single} horizon radius given by (11)%
\begin{equation}
r_{h}=\frac{1}{2}\left[ 1+\sqrt{1+4\Upsilon ^{2}}\right] \geq 1,
\end{equation}%
which does \textit{not} introduce any restriction on $\Upsilon$ unlike in (13). Clearly, the horizon has a greater area than its Schwarzschild counterpart, so that the bulk effects on the brane increase the entropy and decrease the temperature of the black hole, i.e., bulk effects tend to strengthen the gravitational field. This is in direct contrast to the weakening of the field in the brane analogue of Reissner-Nordstr\"{o}m spacetime (11).

DMPR \cite{Dadhich:2000} have offered several arguments as to why the negative sign before $\Upsilon^{2}$ in (14) should be a more logical choice in the braneworld scenario. For instance, the choice is consistent with the confinement to the brane brought about by the negative cosmological constant in the bulk. They have shown that the effective energy density $\mathcal{U}$ on the brane contributed by the free gravitational field in the bulk, which is proportional to $\Upsilon^{2}$, is negative. Also, since the bulk effect projected onto the brane is supposed only to introduce $\left( \frac{1}{r^{2}}\right)$ corrections to the general relativistic Schwarzschild black hole, it is desirable that the nature of the central singularity be preserved under the projection. The singularity in the non-general relativistic metric (14) is \textit{spacelike} thus consistently preserving the original nature of Schwarzschild singularity, while for the RN-like metric (11) the singularity is timelike thereby qualitatively changing the nature of Schwarzschild singularity.

\section{Strong field lensing observables}
\label{sec3}

Light bends in the field of gravity that is the essential physical cause of what we call strong field gravitational lensing in the vicinity of the photon sphere (light bending onto itself) providing images of the source and other observables. Lensing is a powerful diagnostic that can be profitably used in Einstein's as well as in various modified gravity theories as well (see, e.g., \cite{Izmailov:2019}). Bozza \cite{Bozza:2002}, in his pioneering work, derived generic analytical formulas for the strong field relativistic images and their main features in the static spherically symmetric black holes. Here we provide only the outline - the details may be found in his original paper. The idea of strong field lensing differs fundamentally from that usually adopted in the weak field lensing. While the latter takes the deviation from the Minskowski space in the form of PPN expansions in terms of small $\left(\frac{M}{r}\right)$ (see \cite{Keeton:2005}), the strong field limit starts from the complete capture of the photon at the photon sphere, where the bending angle diverges, and takes the leading order of light deflection away from that divergence. We are not going to deal with the strong field lensing in the case of Reissner-Nordstr\"{o}m analogue (11) for which the analyses have already been carried out by Bozza \cite{Bozza:2002} (though $\Upsilon$ is not the electric charge!). We shall focus below on the generic metric (8).

\begin{figure}
	\centering 
	\includegraphics[width=0.4\textwidth, angle=0]{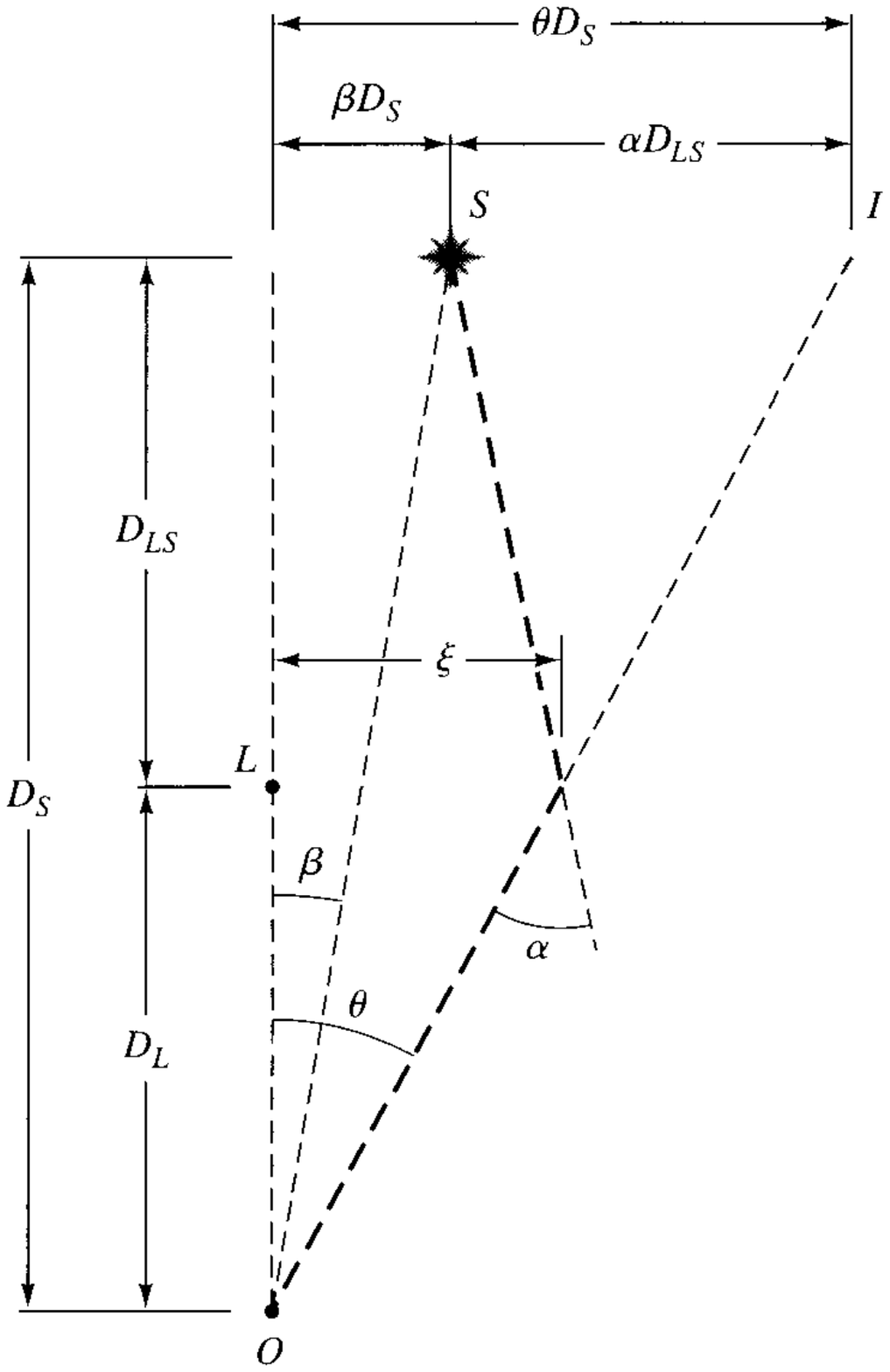}	
	\caption{Lens geometry (Figure taken from J.B. Hartle, \textit{Gravity: An Introduction to Einstein's General Relativity}, Addison-Wesley, San Francisco (2003), p.236} 
	\label{figure-1}%
\end{figure}

\begin{figure}
	\centering 
	\includegraphics[width=0.4\textwidth, angle=0]{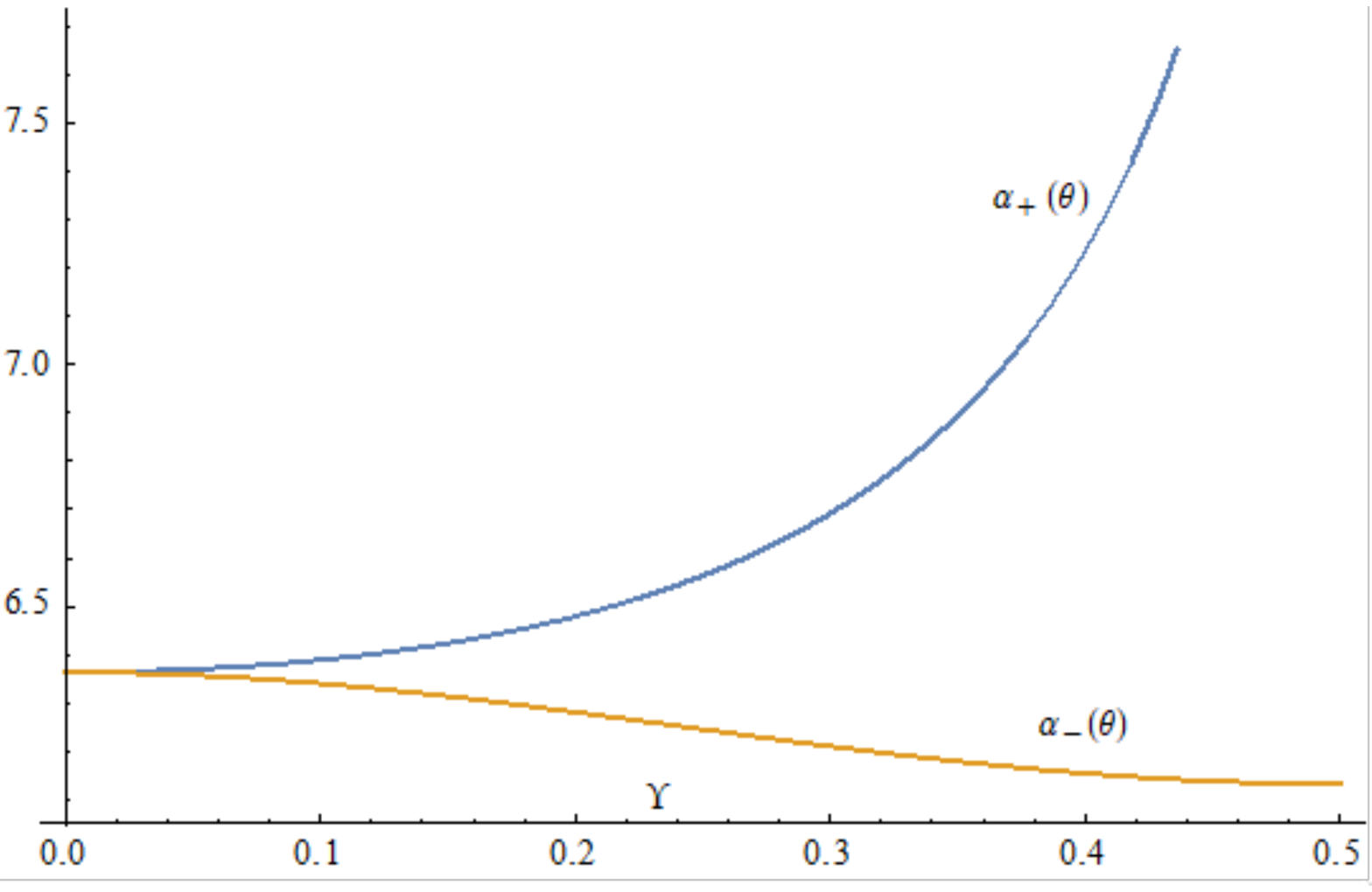}	
	\caption{The angle of light deflection $\alpha _{-}$ by DMPR$-$monotonically decreases with increase in $\Upsilon $ and is less than that by DMPR$+$, which shows that the tidal charge in DMPR$-$ has some kind of a repulsive effect compared to the tidal charge in DMPR$+$, for which $\alpha _{+}$ monotonically increases. The Schwarzschild value occurs at $\Upsilon =0$.} 
	\label{figure-2}%
\end{figure}

Complete capture of photon takes place at photon sphere with its radius $r = r_{m}$ defined by the largest positive root of the equation
\begin{equation}
\frac{A^{\prime }}{A} = \frac{C^{\prime }}{C},
\end{equation}%
where primes denote differentiation with respect to $r$. An incoming photon from infinity is deflected by the gravitating source at a closest approach distance $r_{0}$ corresponding to an impact parameter
$$u=\sqrt{C(r_{0})/A(r_{0})},$$
before emerging on the outgoing side. The photon trajectory is deflected by an angle $\alpha$ obtained directly from the null geodesic $d\tau^{2} = 0$ of the general metric (8) \cite{Virbhadra:2000}: 
\begin{eqnarray}
\alpha (x_{0}) &=&I(x_{0})-\pi , \\
I(x_{0}) &=&\int_{r_{0}}^{\infty }\frac{2\sqrt{B}dr}{\sqrt{C}\sqrt{\frac{C}{%
C_{0}}\frac{A_{0}}{A}-1}},
\end{eqnarray}%
where $C_{0}\equiv C(r_{0})$ etc. When $r_{0}=r_{m}$, the integral (19) diverges logarithmically and the photon is captured. Bozza \cite{Bozza:2002} derived an analytical expansion of the strong field deflection angle close to the divergence in the form
\begin{equation}
\alpha (\theta )=\overline{a}\log \left[ \frac{\theta D_{L}}{u_{m}}-1\right]
+\overline{b}+O(u-u_{m})^{2},
\end{equation}%
where $\overline{a},\overline{b}$ are the coefficients derived from the metric functions and their derivatives at $r_{0}=r_{m}$, $\theta =u/D_{L}$ is the angular separation between the image and the lens and $D_{L}$ is the distance between the observer and the lens (Fig.1) and $u=u_{m}$ corresponding to $r_{0}=r_{m}$ is the shadow radius. The diameter $2u_{m}$ is often called the size of the shadow. The difference in $\alpha_{\pm}(\theta)$ corresponding to DMPR$\pm$ is quite significant, which is displayed in Fig.2.

The quantities involved in the deflection angle $\alpha (\theta)$ for the metric (8) are given by $\overline{a}$, $\overline{b}$, $u_{m}$ \cite{Bozza:2002}
\begin{eqnarray}
\overline{a} &=&\sqrt{\frac{2A_{m}B_{m}}{C_{m}^{\prime \prime}A_{m}-C_{m}A_{m}^{\prime \prime }}} \\
\overline{b} &=&-\pi +b_{R}+\overline{a}\ln \left[ \frac{C_{m}\left(
1-A_{m}\right) ^{2}\left( C_{m}^{\prime \prime }A_{m}-C_{m}A_{m}^{\prime
\prime }\right) }{A_{m}^{3}C_{m}^{\prime 2}}\right]  \\
u_{m} &=&\sqrt{\frac{C_{m}}{A_{m}}},
\end{eqnarray}%
where $b_{R}$ is a definite integral that can in general be computed only numerically. Depending on the metric, e.g., for the Schwarzschild black hole, the integral however can be solved also analytically leading to $b_{R}=2\ln \left[ 6\left( 2-\sqrt{3}\right) \right] =0.9496$.

The strong lensing observables are defined by the angular shadow radius $\theta _{\infty }$, image separation $s$ and magnitude $r$, which are (see, for details, \cite{Bozza:2002}): 
\begin{eqnarray}
\theta _{\infty } &=&\frac{u_{m}}{D_{L}}, \\
s &=&\theta _{\infty }e^{\frac{\overline{b}-2\pi }{\overline{a}}}, \\
r &=&e^{\frac{2\pi }{\overline{a}}}.
\end{eqnarray}%
There are quite significant differences in the above observables for DMPR$\pm $ as displayed in Figs.3,4,5.

\begin{figure}
	\centering 
	\includegraphics[width=0.4\textwidth, angle=0]{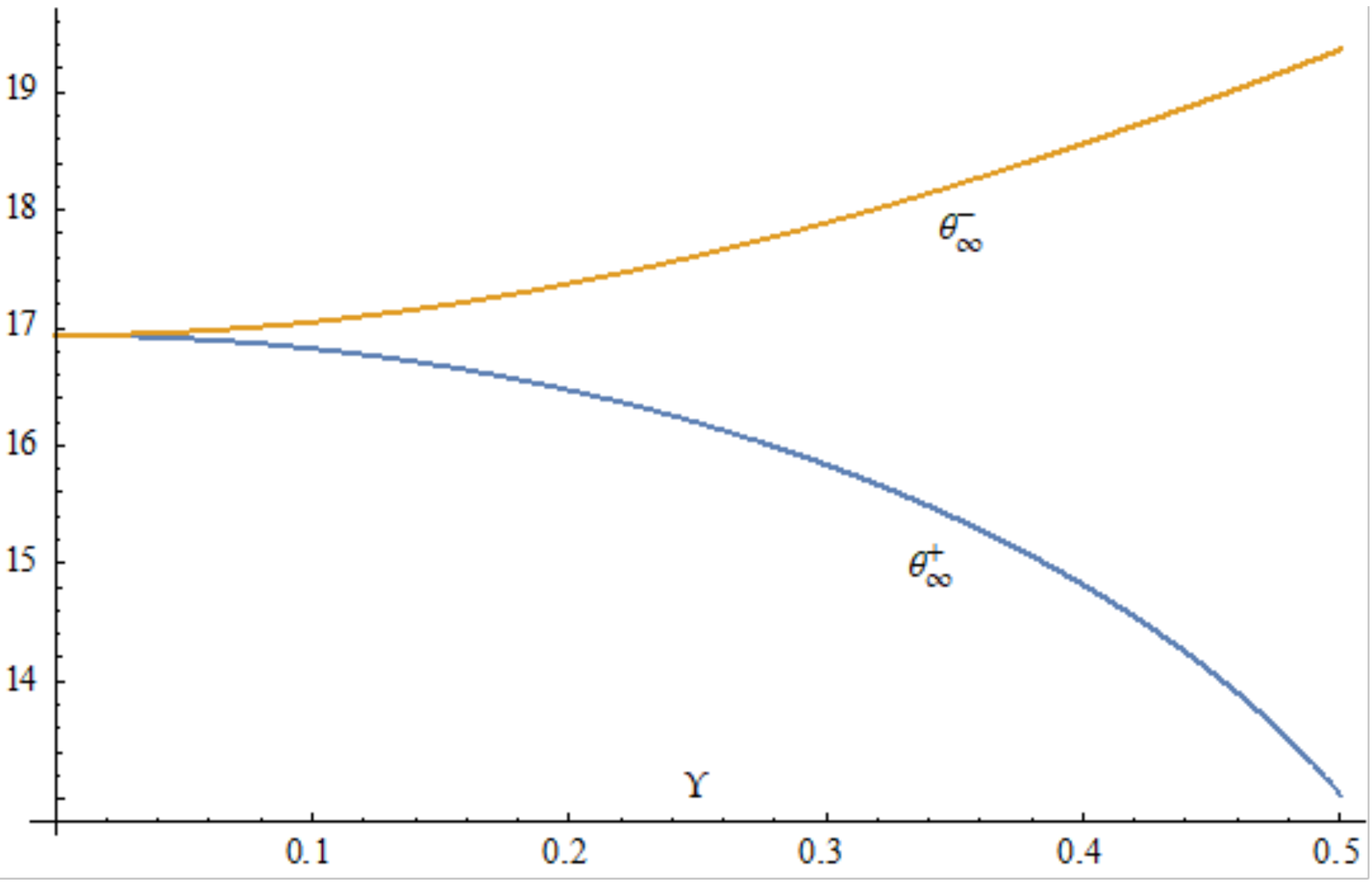}	
	\caption{The angular radius of the shadow $\theta _{\infty }^{-}$ monotonically increases while $\theta _{\infty }^{+}$ monotonically decreases. The benchmark is the Schwarzschild value that occurs at $\Upsilon = 0$, which is $\theta _{\infty }^{\pm }\simeq 17$ $\mu \textmd{arcsec}$.} 
	\label{figure-3}%
\end{figure}

\begin{figure}
	\centering 
	\includegraphics[width=0.4\textwidth, angle=0]{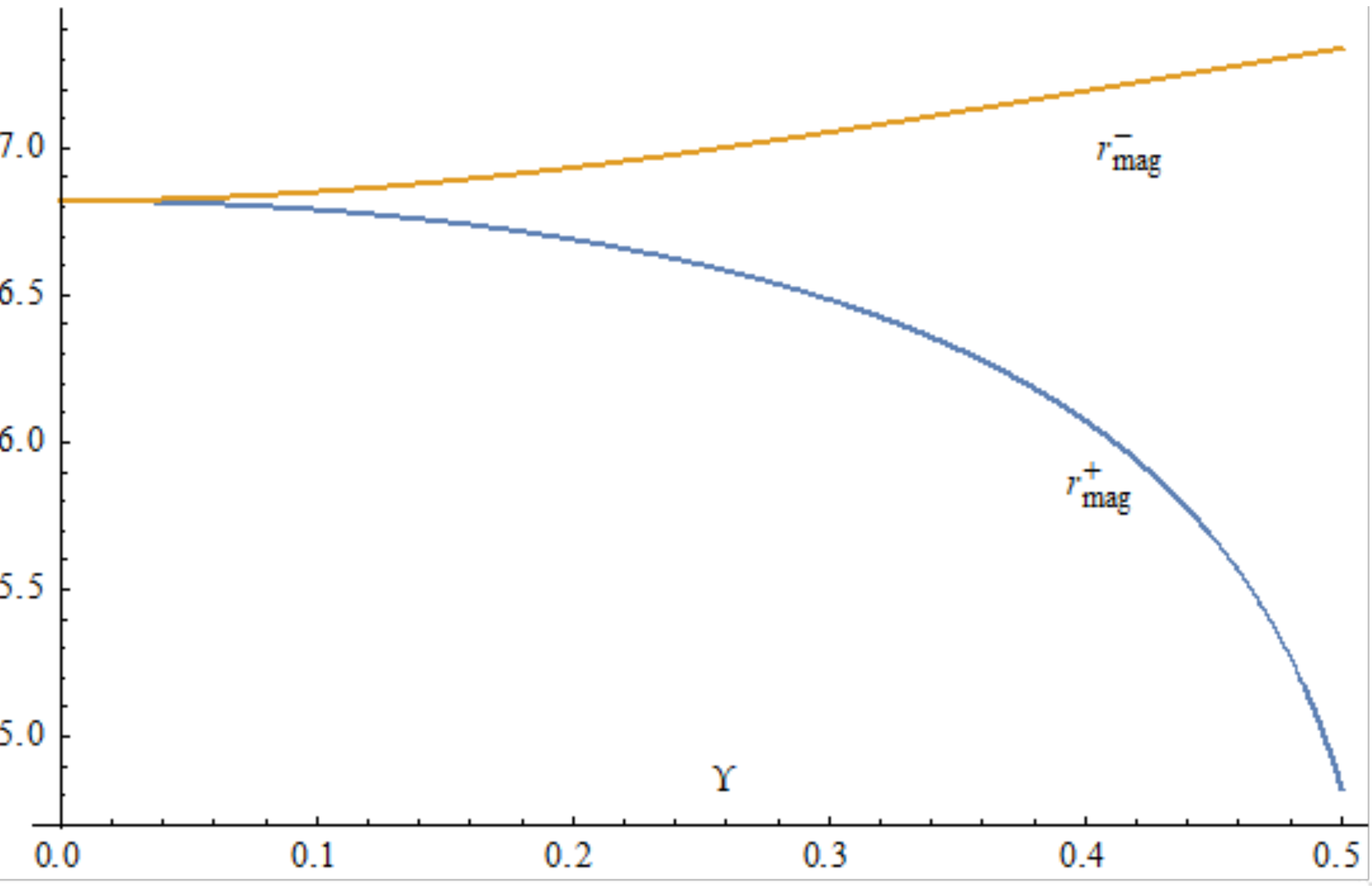}	
	\caption{The logarithm of the magnification ratio $r_{mag}^{\pm}$ in Eq.(31) for DMPR$\pm$. The plots indicate that the image of sources lensed by DMPR$-$ is brighter than the one by DMPR$+$.} 
	\label{figure-4}%
\end{figure}

\begin{figure}
	\centering 
	\includegraphics[width=0.4\textwidth, angle=0]{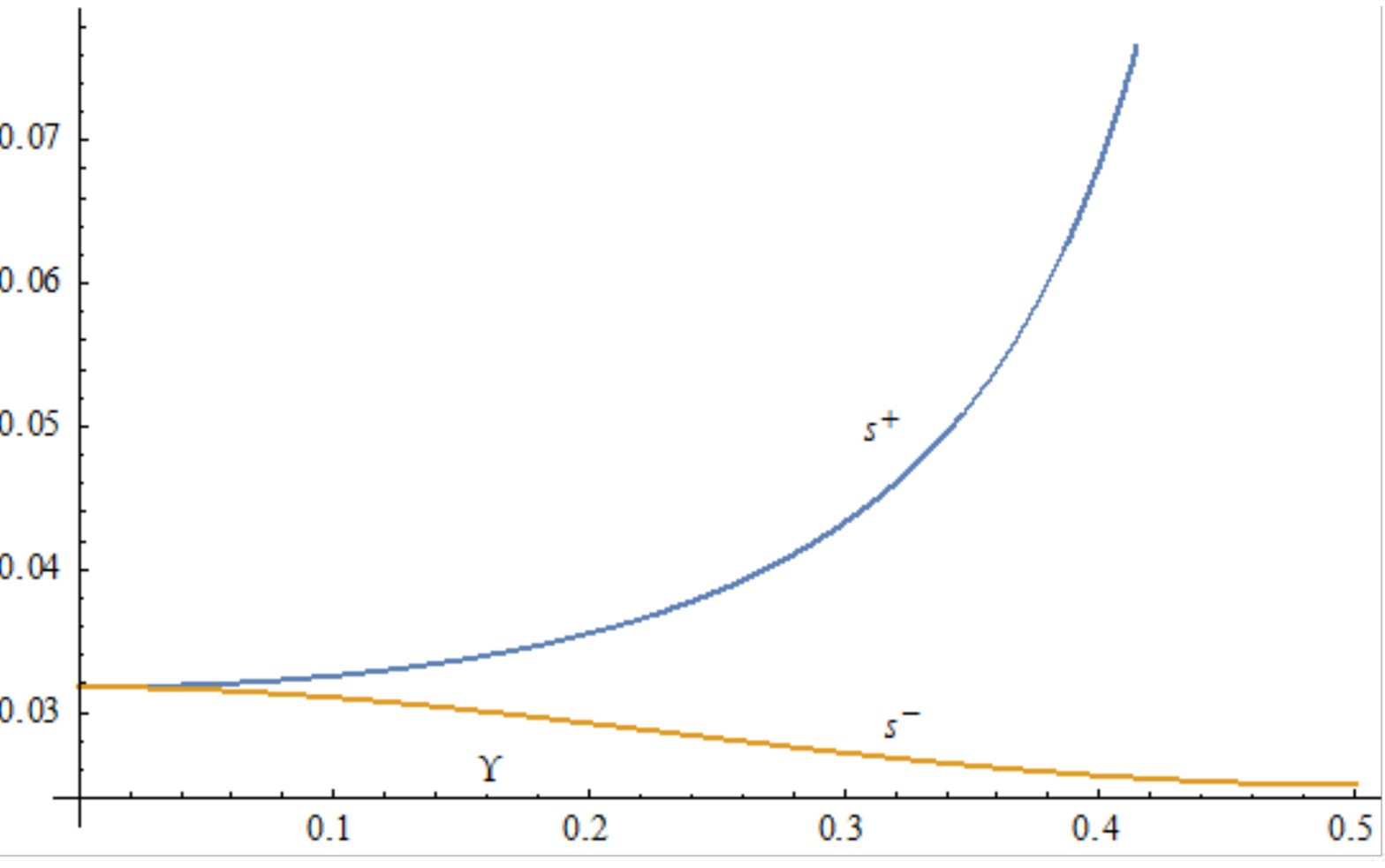}	
	\caption{The image separation $s^{-}$ decreases with increasing $\Upsilon$ showing that the images become gradually unresolvable, contrary to the case of $s^{+}$.} 
	\label{figure-5}%
\end{figure}

\section{QNM modes, Lyapunov and Pretorius-Khurana exponents}
\label{sec4}

Null unstable circular geodesics in all spherically symmetric asymptotically flat black hole spacetime can be used to derive the quasinormal modes (QNM) of the black hole, which describe its characteristic modes of vibration \cite{Hawking:1972,Israel:1967}. In the geometrical optics limit, QNMs can be determined by the parameters of the circular null geodesics. The physical mechanism behind these modes of free vibration can be interpreted in terms of null particles, trapped at the unstable circular geodesics, that leak out slowly to outside observers \cite{Kokkotas:1999,Nollert:1999,Press:1971,Goebel:1972,Ferrari:1984,Mashhoon:1985}. The leaking timescale is given by the Lyapunov exponent, which can be expressed in terms of the second derivative of the effective radial potential for geodesic motion \cite{Cardoso:2009}. The relation between QNMs and circular null geodesics is generic being valid for all asymptotically flat, spherically symmetric black hole spacetimes.

The real part of the complex QNM frequencies is determined by the angular velocity at the unstable null circular geodesic and the imaginary part corresponds to the instability timescale of the orbit. The relation between QNMs and unstable circular null geodesics is quite general, being valid in the eikonal limit for any static, spherically symmetric, asymptotically flat spacetime. The (multiples of) QNM frequencies $\omega_{\scriptsize{\textmd{QNM}}}$ in the WKB approximation by Iyer \cite{Iyer:1987} are
given for large $\ell$ by \footnote{
An important point is to be noted here. We explicitly wrote $2M\sigma$ in the formula (27) since Iyer \cite{Iyer:1987} calculated the QNM frequencies by defining $M\sigma =\omega _{\scriptsize{\textmd{QNM}}}$; his more general WKB formula does yield the the right hand side of (27) for large $\ell$. He took the Schwarzschild metric as $g_{tt} = 1 / g_{rr} = 1 - \frac{2}{r}$. On the other hand, Bozza \cite{Bozza:2002} in his calculation of the lensing observables took the Schwarzschild metric as $g_{tt} = 1 / g_{rr} = 1 - \frac{1}{r}$. Thus, to connect the QNM frequencies of Iyer with the lensing observables of Bozza, as done by Stefanov et al. \cite{Stefanov:2010}, it is necessary to adjust the units: What is $M\sigma $ for Iyer is $2M\sigma$ for Bozza. Hence the factor $\frac{1}{2}$ before the square bracket.}
\begin{equation}
\omega _{\scriptsize{\textmd{QNM}}}=M\sigma =\frac{1}{2}\left[ \Omega _{m}\ell -i\left( n+%
\frac{1}{2}\right) \left\vert \lambda \right\vert \right] 
\end{equation}%
where $n$ is the number of overtones and $\ell$ is the angular momentum of the perturbation. Cardoso et al \cite{Cardoso:2009} showed that the real part of the frequencies is determined by the angular velocity $\Omega_{m}$ of the last circular null geodesic and the parameter $\lambda$ is the Lyapunov exponent determining the inverse of the instability timescale associated with this null geodesic motion. Although the WKB expression is formally valid only in the eikonal limit ($\ell \gg 1$), it can still yield surprisingly accurate predictions even for low values of $\ell$ \cite{Cardoso:2009, Iyer:1987, Berti:2005}.

In a remarkable work, Stefanov et al. \cite{Stefanov:2010} have shown that the QNM modes can be measured by measuring the strong field lensing observables shown in the preceeding section as follows (restoring $c$) 
\begin{eqnarray}
\lambda  &=&\frac{c}{u_{m}\overline{a}}, \\
\overline{a} &=&\frac{2\pi }{\ln \overline{r}}, \\
\Omega _{m} &=&\frac{c}{u_{m}},
\end{eqnarray}%
where $\overline{r}$ is the flux ratio 
\begin{equation}
\overline{r}=\frac{\mu _{1}}{\sum_{n=2}^{\infty }\mu _{n}}
\end{equation}%
in which $\mu _{1}$ and $\mu _{n}$ are the magnifications of the first and $n$th image respectively. Converted to magnitudes, the difference between the first and all other images is defined by $r_{mag}=2.5$Log$\overline{r}$ \cite{Bozza:2002}.

Alternatively, the QNM frequencies can be expressed as \cite{Stefanov:2010} 
\begin{eqnarray}
\lambda &=&\frac{c\ln \overline{r}}{2\pi D_{OL}\theta _{\infty }}, \\
\Omega _{m} &=&\frac{c}{D_{OL}\theta _{\infty }},
\end{eqnarray}%
that allow us to infer $\omega _{\scriptsize{\textmd{QNM}}}$ by measuring the decrease in the brightness $\overline{r}$ of images with increasing $n$ (number of loops), the lens-observer distance $D_{OL}$ and the angular shadow radius $\theta_{\infty}$.

\begin{figure}
	\centering 
	\includegraphics[width=0.4\textwidth, angle=0]{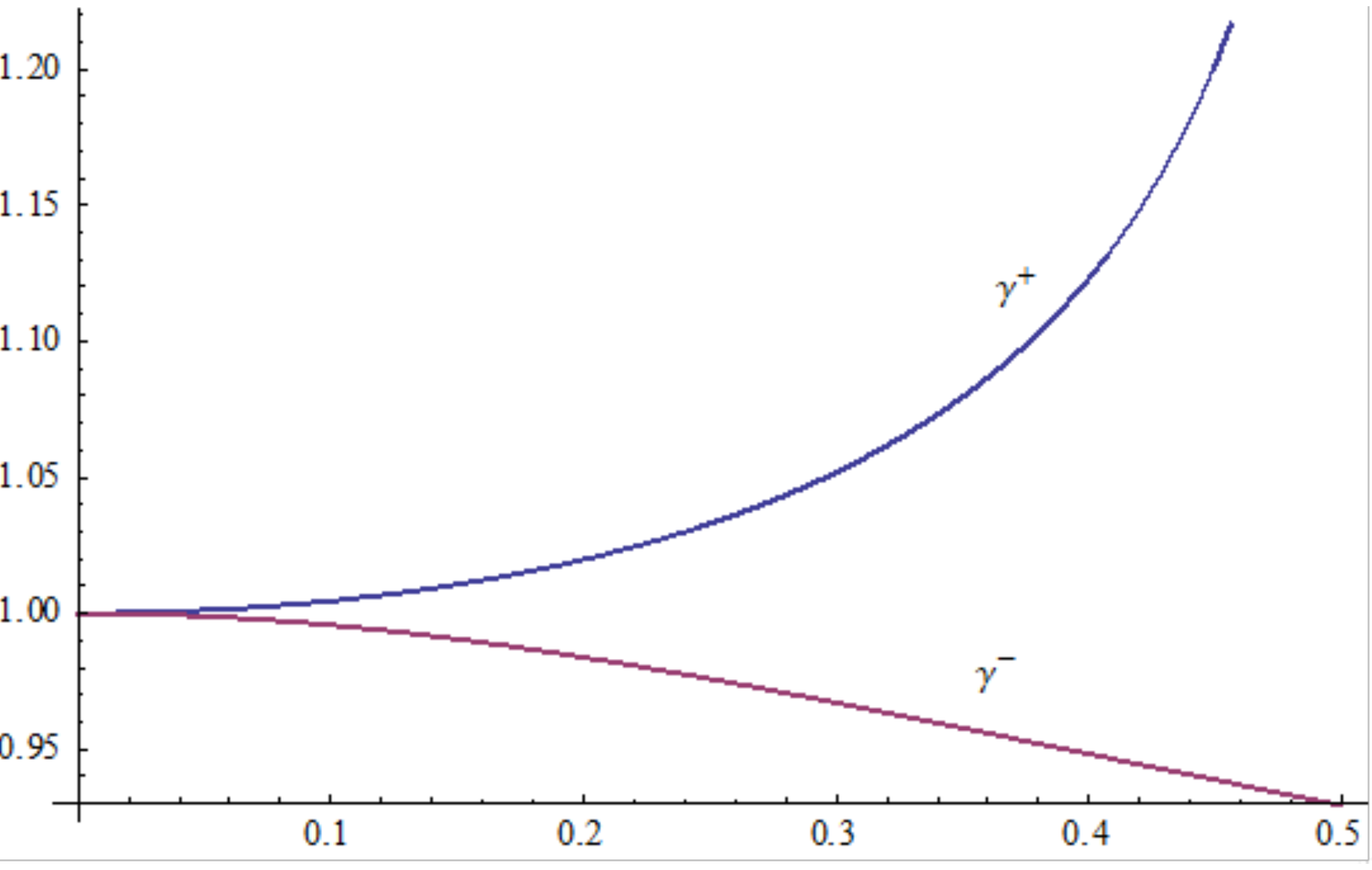}	
	\caption{The Pretorius-Khurana exponent $\gamma ^{-}$ shows that unstable null circular orbits leak out photons more profusely. Consequently, the accretion disk of DMPR$-$ should be more luminous than that of DMPR$+$.} 
	\label{figure-6}%
\end{figure}

\begin{figure}
	\centering 
	\includegraphics[width=0.4\textwidth, angle=0]{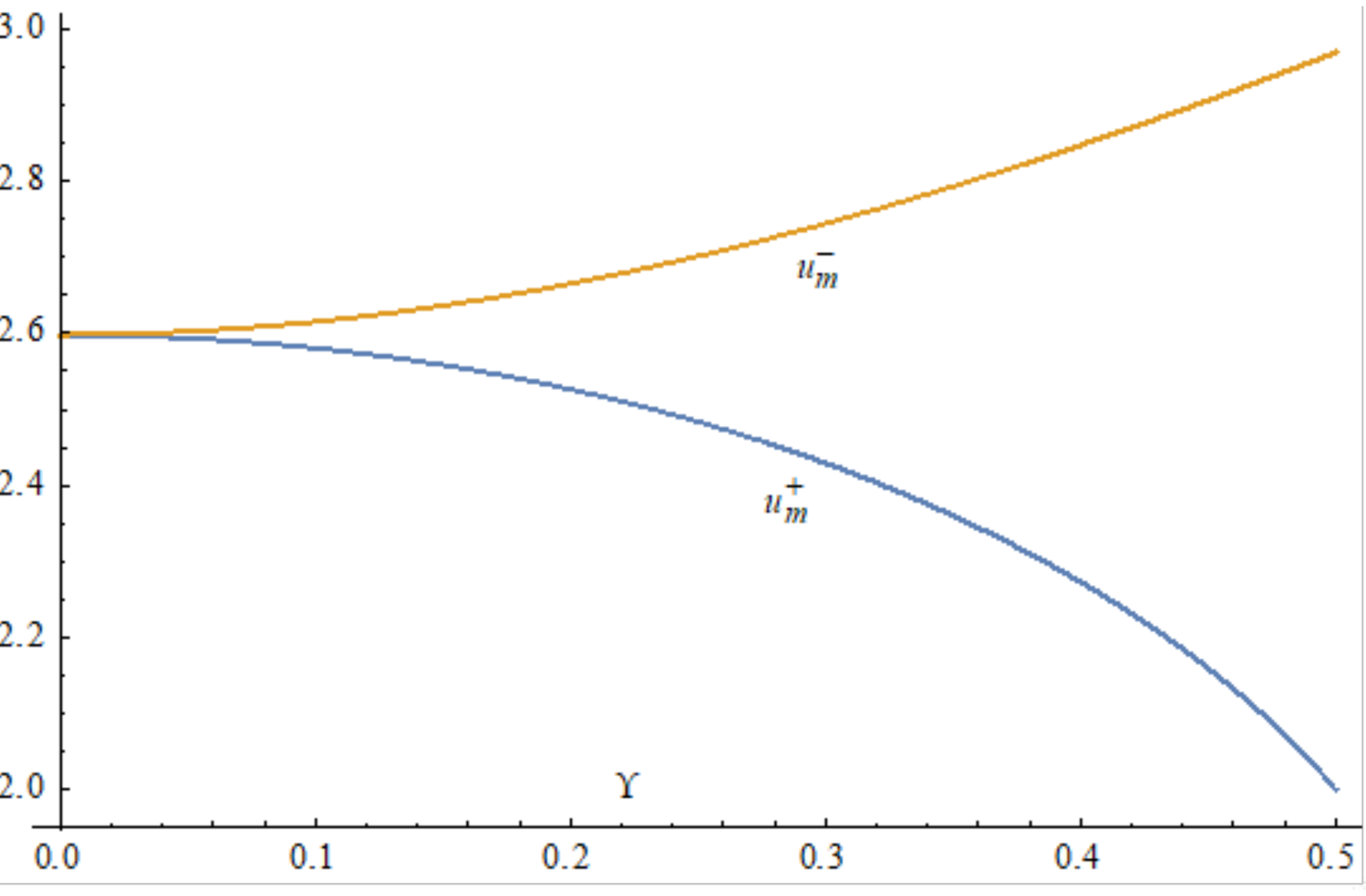}	
	\caption{The plot of $u_{m}^{-}$ is seen to be monotonically increasing from the Schwarzschild value $2.6$, while $u_{m}^{+}$ decreases from that value. This behavior was used in Sec.5 to arrive at the upper limits on $\Upsilon $.} 
	\label{figure-7}%
\end{figure}

\begin{table}
\centering
\begin{tabular}{c c c c} 
 \hline
$\Upsilon$ & $n$ & $\ell$ & $\omega _{\scriptsize{\textmd{QNM}}}(\textmd{DMPR}-)$ \\
 \hline
0 & 0 & 10 & 1.9245-0.0962i   \\ 
0.3 & - & - & 1.8219-0.0941i   \\ 
0.5 & - & - & 1.6837-0.0905i   \\ 
2 & - & - & 0.8939-0.0563i      \\ 
0 & 4 & 30 & 5.7735-0.8660i   \\ 
0.3 & - & - & 5.4657-0.8477i   \\ 
0.5 & - & - & 5.0512-0.8152i   \\ 
2 & - & - & 2.6818-0.5075i     \\ 
0 & 8 & 50 & 9.6225-1.6358i  \\ 
0.3 & - & - & 9.1095-1.6013i  \\ 
0.5 & - & - & 8.4187-1.5398i  \\ 
2 & - & - & 4.4697-0.9587i   \\
 \hline
$\Upsilon \leq 0.5$ & $n$ & $\ell$ & $\omega _{\scriptsize{\textmd{QNM}}}(\textmd{DMPR}+)$ \\
 \hline
0 & 0 & 10 & 1.9245-0.0962i \\
0.3 & - & - & 2.0581-0.0978i \\ 
0.5 & - & - & 2.5-0.0883i  \\
0 & 4 & 30 & 5.7735-0.8660i \\
0.3 & - & - & 6.1745-0.8805i \\
0.5 & - & - & 7.5-0.7954i \\ 
0 & 8 & 50 & 9.6225-1.6358i \\
0.3 & - & - & 10.2908-1.6632i \\
0.5 & - & - & 12.5-1.5026i \\ 
 \hline
\end{tabular}
\caption{The QNM frequencies in the eikonal limit for two types of metrics $d\tau_{\pm }^{2}$ [see Eqs.(11,12) and (14,15)]. For $d\tau _{-}^{2}$, the range of $\Upsilon$ is unrestricted, while for $d\tau _{+}^{2}$ (DMPR$+$), the range of $\Upsilon $ is restricted by $0\leq \Upsilon \leq 0.5$ as demanded by the reality of the horizon radius [Eq.(13)]. Note that the table is only for illustrative purposes - observational limits on $\Upsilon $ are further restricted, see Sec.5. The Schwarzschild black hole QNM frequencies correspond to the rows $\Upsilon =0$. For fixed $n$, $\ell$, the real and imaginary parts decrease with increasing $\Upsilon$ in the spacetime DMPR$-$, while these parts increase with increasing $\Upsilon$ in the spacetime DMPR$+$.}
\label{Table1}
\end{table}

Also the time delay $\Delta T_{2,1}$ between the emergence of the first and the second relativistic images can be expressed as \cite{Stefanov:2010}
\begin{equation}
\Delta T_{2,1}=\frac{2\pi }{\Omega _{m}}
\end{equation}%
and a relation between the Lyapunov instability time%
\begin{equation}
\ln \left( \frac{\mu _{n+1}}{\mu _{n}}\right) =-\frac{2\pi \lambda u_{m}}{c}.
\end{equation}%
The Pretorius-Khurana exponent $\gamma$ is defined by \cite{Cardoso:2009, Pretorius:2007}
\begin{equation}
\gamma =\frac{\Omega _{m}}{2\pi \lambda }=\frac{\overline{a}}{2\pi }.
\end{equation}%
The exponent can thus be defined as the ratio of typical orbital time scale $T_{\Omega} = \frac{2\pi}{\Omega _{m}}$ and an instability time scale $T_{\lambda} = \frac{2\pi}{\lambda}$ \cite{Cardoso:2009}. Small values of $\gamma$ correspond to a strong Lyapunov instability, i.e., more photons from the unstable circular null orbits tend to leak out to outside observers, hence the accretion disks would appear brighter. The comparative behavior of $\gamma ^{\pm }$ is shown in Fig.6.

Following Bozza's method for the metrics (11) and (14), the basic quantities $u_{m}$, $\overline{b}$ and $\overline{a}$, through which other observables are defined, are given for the two metrics by%
\begin{eqnarray}
u_{m}^{\pm } &=&\left( \frac{1}{4\sqrt{2}}\right) \frac{\left( 3+\sqrt{9\pm
32\Upsilon ^{2}}\right) ^{2}}{\sqrt{3\pm 8\Upsilon ^{2}+\sqrt{9\pm
32\Upsilon ^{2}}}}, \\
\overline{a}_{\pm } &=&\frac{r_{m}^{\pm }\sqrt{r_{m}^{\pm }\pm 2\Upsilon ^{2}%
}}{\sqrt{(3-r_{m}^{\pm })r_{m}^{\pm 2}\pm 9\Upsilon ^{2}r_{m}^{\pm
}+8\Upsilon ^{4}}}, \\
\overline{b}_{\pm } &=&-\pi +0.9496-1.5939+\overline{a}_{\pm }\left[ 2\ln
\left( r_{m}^{\pm }\pm \Upsilon ^{2}\right) ^{2}X_{\pm }\right]  \\
X_{\pm } &=&\frac{\left( 3-r_{m}^{\pm }\right) r_{m}^{\pm 2}\pm 9\Upsilon
^{2}r_{m}^{\pm }+8\Upsilon ^{4}}{\left( r_{m}^{\pm }\pm 2\Upsilon
^{2}\right) ^{3}\left( r_{m}^{\pm 2}-r_{m}^{\pm }\pm \Upsilon ^{2}\right) }
\\
r_{m}^{\pm } &=&\left( \frac{3}{4}\right) \left( 1+\sqrt{1\pm \frac{%
32\Upsilon ^{2}}{9}}\right) ,
\end{eqnarray}%
where $\pm $ refer to the functions calculated for the metrics $d\tau_{\pm}^{2}$. Correspondingly, we can calculate the observables $\theta _{\infty}^{\pm }$, $s^{\pm },r^{\pm },\gamma ^{\pm }$ and $\omega _{\scriptsize{\textmd{QNM}}}$ from Eqs.(24), (25), (26) and (36).

\section{Observational limits on $\Upsilon$ for SgrA*}
\label{sec5}

The tidal charge modification to the magnitudes of different strong field observables away from those of the Schwarzschild ($\Upsilon =0$) for both the metrics (11) and (14) are evident from the Figs.(2-7) wherein we had displayed the variations of $\theta_{\infty}^{\pm}$, $s^{\pm}$, $r^{\pm}$, $\lambda ^{\pm } u_{m}^{\pm}$, $\gamma ^{\pm}$ with respect to $\Upsilon$. It is quite clear that accurate measurements, when available, should put limits on the magnitudes of the tidal charge $\Upsilon$. The Schwarzschild value that occurs at $\Upsilon = 0$ serves as the benchmark for all observables. A prime candidate for this purpose is the measurement of the shadow diameter%
\footnote{%
More precise terminology would be shadow size rather than the shadow diameter because the image is blurred due to many physical factors, e.g., scattering of light by the surrounding plasma medium. However, in the present context, we adhere to the term shadow diameter to which an exact Schwarzschild value can be attributed and deviations from it is assumed to be caused by the tidal charge.}of the supermassive black hole SgrA* living at the center of our galaxy Milky Way \cite{Boehle:2016}. Analytical studies \cite{Psaltis:2015} predict the angular shadow radius $\theta _{\infty}$ to lie in the range $(5\pm 0.2)\Theta _{M}$, where $\Theta _{M}=GM/D_{L}c^{2}$ $\approx 5 \mu as$, using mass $M=4\times 10^{6}M_{\odot }$ and the distance from Earth $D_{L}=8.1$ kpc \cite{Psaltis:2015}. The uncertainty in radius arises from the variation of the spin and inclination. Thus the angular shadow diameter becomes $2\theta_{\infty }=2\times (5\pm 0.2)\Theta_{M}\approx (50\pm 2) \mu as$.

Later quantitative study predicted the angular shadow diameter and asymmetry to be $\approx 50$ $\mu as$ and $\lesssim 3 \mu as$, respectively \cite{Zhu:2019}. It was suggested therein that if the measured values depart from these values, it could indicate a violation of the \textquotedblleft no-hair theorem" \cite{Hawking:1972, Israel:1967, Israel:1968, Carter:1971, Robinson:1975}. However, the measurements are complicated by many factors. Refractive scattering of the emerging light by the turbulent ionized medium distorts the image of Sgr A* affecting its apparent size, which include effects like image wander, distortion, and asymmetry. If the turbulence in the scattering medium of Sgr A* is assumed to have a Kolmogorov slope \cite{Beresnyak:2011, Akhmetev:2022}, then the refractive image wander, distortion, and asymmetry leads to corrections of magnitudes $0.53,0.72$ and $0.52$ $\mu as$ at $230$ GHz \cite{Psaltis:2015}. It will turn out that these values are much less than the correction due to the tidal charge obtained from observations.

The first Event Horizon Telescope (EHT) observations using a global interferometric array of eight telescopes operating at a wavelength of $1.3$ mm was conducted in 2017 \cite{Akiyama:2022}. The data resolved a compact emission region. A variety of imaging and modeling analyses all support an image that is dominated by a bright, thick ring with an observed angular
shadow diameter of $2\theta _{\infty}^{\scriptsize{\textmd{obs}}}=\left(51.8\pm 2.3\right) \mu as$ at $68\%$ confidence level. The ring has a modest azimuthal brightness asymmetry and a comparatively dim interior. Using numerical simulations, it was demonstrated that the EHT images of Sgr A* are consistent with the expected appearance of a Kerr black hole with mass $\sim 
$ $4\times 10^{6}M_{\odot }$. The effect of SgrA* spin ($a/M=0.44$ at $1\sigma $ level \cite{Kato:2010}) introduces only a modest asymmetry ($\pm 1$ $\mu as$) in the shadow which is very nearly circular for low spin \cite{Johannsen:2010}, so we ignore spin and given the uncertainties in the observed mass and distance, we heuristically try to find an upper limit on $\Upsilon$ on the basis of the observed angular shadow diameter. The Schwarzschild value for the angular shadow diameter (which is twice the minimum impact parameter $u_{m}$) for the above values of mass and distance, is $2\theta _{\infty}^{\scriptsize{\textmd{Sch}}} = \left. \frac{2u_{m}^{\scriptsize{\textmd{Sch}}}}{D_{L}}\right\vert _{\Upsilon =0}=50.783$ $\mu as$. Now compare the behavior of the plots of $u_{m}^{-}$ and $u_{m}^{+}$ (Fig.7) in units of $2M=1$; the plot of $u_{m}^{-}$ is seen to be monotonically increasing from the Schwarzschild value $u_{m}^{\scriptsize{\textmd{Sch}}} = \frac{3\sqrt{3}}{2}\times 2M=2.598$ or $2\theta _{\infty}^{\scriptsize{\textmd{Sch}}}=50.783$ $\mu as$, while that of $u_{m}^{+}$ monotonically decreasing from $50.783$ $\mu as$. On the other hand, we note from the observational uncertainties that \cite{Akiyama:2022} $49.5<2\theta_{\infty }^{\scriptsize{\textmd{obs}}}<54.1$ $\mu as$, the interval containing $2\theta _{\infty }^{\scriptsize{\textmd{Sch}}}$. If we consider the metric (11) allowing a double horizon, then we have to consider the decreasing $u_{m}^{+}$ or $2\theta _{\infty }^{+}=\left. \frac{2u_{m}^{+}}{D_{L}}\right\vert $. Then the inequality $49.5-2\theta _{\infty }^{+}\leq 0$ holds if $0\leq \Upsilon \leq 0.19178$. If we consider the increasing $u_{m}^{-}$, then the corresponding inequality $54.1-2\theta_{\infty}^{-}\geq 0$ yields $0\leq \Upsilon \leq 0.32467$. The seemingly overlapping intervals do not mean it is physically the same $\Upsilon$ since the metrics are fundamentally different - metric (11) is general relativistic (\textit{sans} the attribute of electric charge) with double horizon, while (14) is not. Therefore the intervals are essentially disconnected. We here advocate the latter interval to be more representative of the braneworld scenario on the ground that, as argued in Sec.2, the extra-dimensional power law correction is \textit{not} expected to change the single Schwarzschild horizon into a double one (of the DMPR$+$) or change its character of spacelike singularity.

\section{Conclusions}
\label{concl}

The QNM frequencies for RN black hole with electric charge studied in \cite{Kokkotas:1988} provide a fundamental source of values. RN black holes are geometrically similar to DMPR$+$ but physically distinct from each other in that $\Upsilon$ in the latter is not the electric but the tidal charge modifying the Schwarzschild black hole. On the other hand, DMPR$-$ black holes are geometrically distinct from DMPR$+.$ One of our aim then was to study their distinctive QNM frequencies in the eikonal limit by comparing the corresponding values from \cite{Kokkotas:1988} applicable for RN (DMPR$+$ or RN-type). The other goal was to study a number of strong field lensing observables (due to Bozza \cite{Bozza:2002}) $\alpha_{\pm}$, $\theta_{\infty}^{\pm}$, $s^{\pm}$, $r^{\pm}$, $\gamma ^{\pm}$, $u_{m}^{\pm }$ including the observed shadow size $2\theta _{\infty}^{\scriptsize{\textmd{obs}}}$ that can constrain $\Upsilon$ from the EHT observation. (Figs. 2-7) display the effect of the extra fifth dimension that significantly differ from those of the pure Schwarzschild black hole ($\Upsilon = 0$), and also between the power-law modified (by the tidal charge $\Upsilon >0$) DMPR$\pm$. All these studies have been done here for the first time, to our knowledge.

We assumed the SgrA* as an illustrative candidate for the black hole that is residing at our Galactic center. The tidal charge effects representing imprints from the extra dimension can potentially be measured from Earth with precision technology in the future that could provide a crucial astrophysical test of Randall-Sundrum braneworld model.

The following are our specific new results, which reveal significant observable differences among the Schwarzschild and DMPR$\pm$ black holes:

$\bullet$ We have exploited the remarkable connection between lensing observables and QNM frequencies discovered in \cite{Stefanov:2010}. Quantitative differences appear in the behavior of observable QNMs of DMPR$\pm$ black holes indicating presence of tidal charge in the spectrum as shown in Table 1. For fixed $n$, $\ell$, the real and imaginary parts of $\omega _{\scriptsize{\textmd{QNM}}}$ decrease with increasing $\Upsilon$ in the spacetime DMPR$-$, while in stark contrast these parts increase with increasing $\Upsilon$ in the spacetime of DMPR$+$ (Table 1).

$\bullet$ We discover a stronger Lyapunov instability of null circular geodesics leaking out QNM frequencies than those around the pure Schwarzschild black hole and DMPR$-$. The reason is that the Pretorius-Khurana exponent $\gamma $ is found to be smaller than that of the Schwarzschild black hole (Fig.6). Thus DMPR$-$ would appear brighter that DMPR$+$ and Schwarzschild black hole.

$\bullet $The best observable is the shadow size $2\theta_{\infty}^{\scriptsize{\textmd{obs}}}$ of SgrA*. Other effects such as the Kolmogorov MHD turbulence around SgrA* do not significantly distort the shadow. Therefore, we have argued that the tidal charge can be constrained by the current Event Horizon Telescope (EHT) observations by $0\leq \Upsilon \leq 0.32467.$

These constraints may not be watertight as it is known that SgrA* has a spin $a_{\ast }=a/M=0.44\pm 0.08$ ($1\sigma $ uncertainty) \cite{Kato:2010}. Thus, a more appropriate analyses should involve spin but EHT observations indicate that the effect of spin on the shadow is $\leq 1\mu as$ \cite{Johannsen:2010}, hence the spin does not significantly alter the obtained constraint on $\Upsilon $. The contour of the shadow of the spinning braneworld black hole with tidal charge has been investigated in Refs.\cite{Schee:2009, Amarilla:2012}. It would be our future task to see how the limit could be sharpened taking into account spin and other distorting effects on the shadow radius.

We shall finally make three clarifying comments:

Most objects in the universe are spinning, so introducing spin in the study of observables would describe a more realistic situation, which is possible using the generic framework of spinning black holes in general relativity or in other theories. There are several important studies on shadow and gravitational lensing by spinning black holes including the study of geodesics around them, see, e.g., \cite{Bozza:2003, Hsiao:2020, Liu:2017, Jiang:2018}. Xavier et al \cite{Xavier:2020} have recently compared the shadows of the Kerr-Newman (KN) and Kerr-Sen (KS) stringy black holes. They found that stringy black hole always has a larger shadow for the same physical parameters and observation conditions. This conclusion in principle accords with the conclusion in the static case studied in our paper, viz., Randall-Sundrum stringy black hole's angular shadow becomes monotonically larger with increasing tidal charge $\Upsilon$ compared to that of the Schwarzschild black hole ($\Upsilon = 0$) (see Fig.3). Gravitational lensing and frame dragging of light in the KN and in the KN-(anti) de Sitter black hole spacetimes have been studied by Kraniotis \cite{Kraniotis:2014}. A more recent study on strong gravitational lensing by Kerr and KN black holes appear in the work of Hsieh et al \cite{Hsieh:2021}, where observability of the relativistic images influenced by the spin $a$ and electric charge $Q$ of the black holes have been thoroughly discussed. In principle, the overall lensing behavior accords well with those in the static case as evidenced in Fig.5 of our paper, e.g., the increase in image separation with increase in $\Upsilon$. When $a=0$, one recovers the results similar to that presented for DMPR$+$. Black hole shadow in KN and spinning regular (singularity-free) black holes has been studied by Tsukamoto \cite{Tsukamoto:2018}. In our future work, we shall consider the spinning DMPR$\pm$ and study the effect of spin $a$ and tidal charge $\Upsilon$ on lensing observables.

2. Note that the branch DMPR$+$ is only \textit{formally} similar to the RNBH [Eq.(8)] but is physically very different from it. First, the field equations are not the traditional Einstein-Maxwell equations since the trace-free $\mathcal{E}_{\mu\nu}$ is a projection of the bulk Weyl tensor on the brane and not the usual Maxwell field tensor $F_{\mu\nu}$. Second, the change $\Upsilon \rightarrow i\Upsilon$ leading to the DMPR$-$ has profound consequences. For instance, DMPR$-$ has completely different horizon properties [compare Eqs.(13) and (16)], shadow and lensing properties as exposed in this paper. For instance, figures show that, while the angular shadow radius $\theta _{\infty}^{+}$ decreases, $\theta _{\infty }^{-}$ increases with increase in $\Upsilon$. This apart, there are many other differences between DMPR$+$ (or formally RNBH) and DMPR$-$, all enumerated throughout the text. In fact, parametric complexification like $\Upsilon \rightarrow i\Upsilon$ is reminiscent of Wick rotation that are powerful enough even to remove, otherwise irremovable, spacetime singularities as shown previously in a different context (see e.g., \cite{Nandi:2008}).

3. A clear roadmap for experimental verification of the theoretical predictions is very important for the present study to be observationally meaningful. Measuring the shadow size ($2u_{m}$) of SgrA* is as yet the best target achieved by the Event Horizon Telescope (EHT), which is a global network of synchronized radio observatories that act as an Earth-sized telescope. It can observe features with an angular resolution of less than $20$ $\mu as$. This is already a tremendous feat achieved by the EHT. This nothwithstanding, the structure of the persistent ring feature in EHT images of Sgr A* is currently not sufficiently resolved to unambiguously confirm the presence of a \textquotedblleft photon ring\textquotedblright\ defined as follows: Photon sphere with radius $r_{m}$ [Eq.(41)] marks the limiting boundary of the strong field at which photons get captured [see Eq.(20)]. Thus, incoming photons from the source undergo $n$ loops in the close vicinity of the photon sphere before escaping to infinity. The weakly-lensed (no loop) direct image, $n=0$, is therefore accompanied by extremely lensed subrings $n=1,2,...$ formed by photons that looped $n$ times around the photon sphere before being intercepted at the observer. The photon ring is the sum of the $n\geq 1$ subrings. Due to the exponential demagnification of subsequent rings, the majority of the photon ring flux will lie in the $n=1$ subring containing information on near-horizon black hole physics.

Thus, the technical challenge is to precisely image the $n=1$ subring, particularly its thickness. The prospects of observing this subring in the next decade by a next-generation Event Horizon Telescope (ngEHT) are quite promising. The project involves designing an array to
substantially enhance the observational capabilities of the EHT. The timeline for ngEHT is that it will add up to $\sim 10$ additional sites worldwide by $\sim 2030$. For an excellent review of the
project, see \cite{Ayzenberg:2023}. For other related recent studies, see \cite{Wielgus:2021, Palumbo:2023, Vagnozzi:2023}.

\section*{Acknowledgments}

We thank three anonymous reviewers for their useful suggestions that led us to make the above clarifying comments. This work was supported by the Russian Science Foundation under grant no. 23-22-00391, https://rscf.ru/en/project/23-22-00391/.

\end{document}